Title: Vehicle Communication using Hash Chain-based Secure Cluster

Authors: Na-Young Ahn(humble@korea.ac.kr), and *__Dong Hoon Lee__ (donghlee@korea.ac.kr)

Affiliation: Graduate School of Inforamtion Security, Korea University, Seoul, Korea

*Corresponding Author

Title: Vehicle Communication using Hash Chain-based Secure Cluster

Authors: Na-Young Ahn(humble@korea.ac.kr), and ***Dong Hoon Lee** (donghlee@korea.ac.kr)

Affiliation: Graduate School of Inforamtion Security, Korea University, Seoul, Korea

*Corresponding Author

# Vehicle Communication using Hash Chain-based Secure Cluster

**Abstract**

We introduce a hash chain-based secure cluster. Here, secure cluster refers to a set of vehicles having vehicular secrecy capacity of more than a reference value. Since vehicle communication is performed in such a secure cluster, basically secure vehicle communication can be expected. Secure hash clusters can also be expected by sharing hash chains derived from vehicle identification numbers. We are also convinced that our paper is essential for future autonomous vehicles by providing secure clustering services using MEC. In the near term, autonomous driving, our paper makes it possible to expect strong and practically safe vehicle communications.

**Introduce**

Recently, the proliferation of quantum computers has increased the possibility of threats of eavesdropping or tampering with data transmitted and received in wireless communications [1-3]. Naturally, these quantum computers will also be great threats to vehicle radio communications. So there has been a lot of research on physical layer security against the threats of such eavesdropping/tampering [4]. Researches to solve this problem are proceeding in two directions. The first is to use quantum cryptography to fundamentally block threats from quantum computers. Next, we will protect the wireless communication channel through physical layer security. At present, the research on the physical layer security is considered more practical. In the near future, autonomous driving is likely to be realized on the road. Autonomous driving basically assumes wireless communication between vehicles. For this reason, physical layer security will be an important issue in vehicle communication in the autonomous driving era.

**Vehicular Secrecy Capacity**

Physical layer security can be considered as a concept of allocating any one of a plurality of wireless channels to be radiated. That is, physical layer security will make very few radio signals to be transmitted to the eavesdroppers, or signals transmitted to the eavesdroppers will transmit signals that are completely different from those to which the legitimate sender is sent. As a result, this physical layer security ensures that a unique wireless channel is allocated between the sender and the receiver. Physical layer security can be considered as a concept of encapsulating any one of a plurality of radiating wireless channels, referrig to Fig.1. That is, according to physical layer security, there is almost no radio signal transmitted to the eavesdropper, or a signal completely different from the signal transmitted to the eavesdropper is transmitted to the legitimate sender. As a result, it can be appreciated that the role of such physical layer security is to allocate a unique wireless channel between the sender and the receiver.

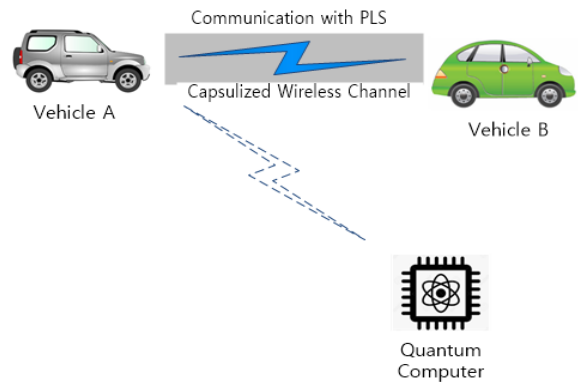

Figure 1: Secure V2V communication with PLS.

In vehicle communication, for safe communication, the ophthalmologist proposed a vehicle communication using a secrecy capacity [5]. In general, the secrecy capacity is defined as the target channel capacity minus the eavesdropping channel capacity. The physical meaning of this secrecy capacity can be thought of as a measure of how well data is transmitted to the target channel without the threat of eavesdroppers. However, since vehicle communication in the real world has no knowledge of the existence of the eavesdropper and the number of eavesdroppers, the definition of such secrecy capacity is not valid. Thus, we define a vehicle secrecy capacity for vehicle communication.

The secrecy capacity for the vehicle is defined as the channel capacity between the target vehicles minus the average channel capacity between the remaining vehicles, referring to Fig. 2. Meaning can be thought of as a figure of how well data is transmitted to a target vehicle.

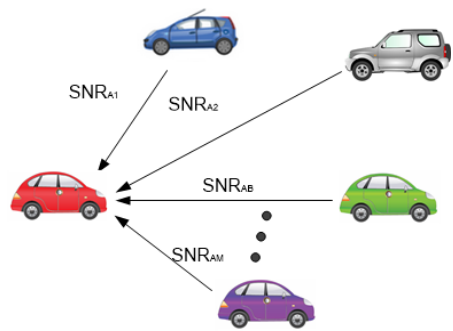

Figure 2: Definition of Vehicular Secrecy Capacity.

We propose a similar secrecy capacity, vehicular secrecy capacity (VSC), which is independent of the existence of an eavesdropper. Generally, the eavesdropper may or may not transmit response messages corresponding to its channel information in response to a communication-initiated signal from the host vehicle. The response message includes channel information that goes through channel states, and this channel information may contain SNR values. A host vehicle may receive channel information from either eavesdroppers or legitimate vehicles. The VSC value is defined under the assumption that the SNR value of the eavesdropper is lower than the average SNR value. Using the SNR values of the received channel information, host vehicles can define the VSC as follows:

$$\text{VSC} = \log_2(1 + \text{SNR}_{AB}) - \log_2(1 + \text{SNR}_{XOR}), \qquad (1)$$

where $\text{SNR}_{XOR} = \frac{\sum_{i=1}^{M} \text{SNR}_{Ai}}{M}$, and M is a number of channel signaling received during the unit time. B represents the target vehicle for communication, and i is the vehicle number, excluding the host vehicle.

**Secure Cluster**

We define a secure cluster for vehicle communication. Here, the secure cluster is a set of vehicles whose vehicle secrecy capacity is greater than or equal to the reference value. Each vehicle can calculate its own secrecy capacity for the vehicle. Each vehicle may propagate to surrounding vehicles via broadcast signals that the calculated secrecy capacity is above a reference value. Referring to Fig. 3, the broadcast signal is a next generation broadcast signal (eg. ATSC 3.0), and may be transmitted by multiplexing an emergency broadcast signal for transmitting an emergency message and a normal broadcast signal for transmitting data for driving a vehicle [6]. The core level signal is an emergency message and a signal for transmitting information transmitted when a vehicle event occurs. The enhancement level signal is a signal for conveying information for driving the vehicle. Such information may typically include driving information of the vehicle.

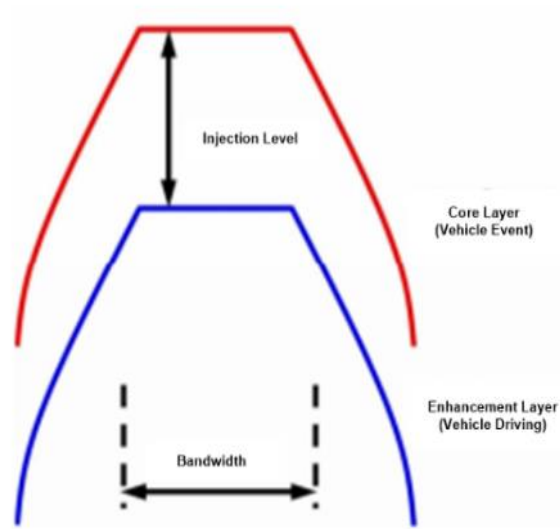

Figure 3: V2V signal using ATSC 3.0 signal.

**Vehicle Hash Chain**

In general, each vehicle has an identification number to identify the vehicle. We formed a hash chain based on the vehicle identification number. Each vehicle stores this vehicle hash chain. Each vehicle may disclose the vehicle hash chain value and the number m corresponding to the last hash operation to other vehicles when the vehicle secrecy capacity is equal to or greater than the vehicle secrecy capacity reference value for vehicle communication, referring to Fig.4.

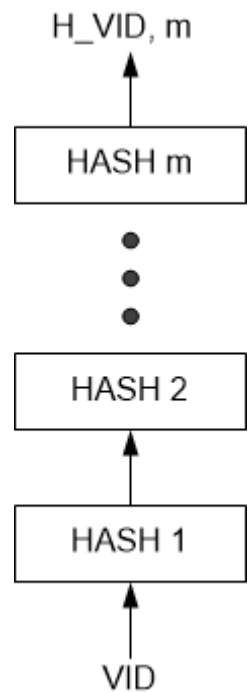

Figure 4: Generation of Vehicle Hash Chain.

**Verification of Vehicle Hash Chain**

Any vehicle that has received a vehicle hash chain value of another vehicle can verify at any time whether the received value has really been transferred from a legitimate vehicle. To this end, each vehicle can verify whether the published vehicle hash chain value is valid by determining whether the published vehicle hash chain value and the hash value performed m times from the vehicle identification number based on the hash algorithm are the same.

**Formatting for Secure Cluster**

Anyone whose vehicle secrecy capacity is greater than or equal to the reference value may disclose a number corresponding to his vehicle hash chain value and the repeated hash operation number. The vehicle creating the secure cluster may add the block to the secure cluster by receiving the number corresponding to this vehicle hash chain value, referring to Fig. 5.

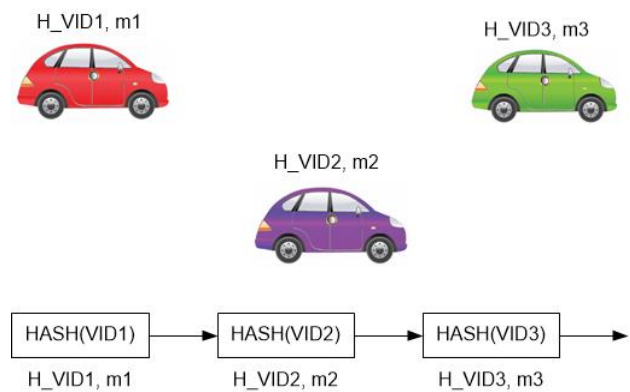

Figure 5: Formation of Secure Cluster.

The hash chain information corresponding to the generated secure cluster will be sent to all vehicles forming the secure cluster. This will allow vehicles with secure clusters to share hash chain information. This hash chain information can be used for data transmission or reception as a kind of shared key.

**Data Transmitting**

The transmitting vehicle may transmit a broadcast signal having data encrypted with its vehicle hash chain value and hash chain information to a plurality of vehicles.

**Data Receiving**

The receiving vehicle receives the broadcast signal from the transmitting vehicle and checks whether the vehicle hash chain value is included in the stored hash chain information. If the vehicle hash chain value is included in the stored hash chain information, the receiving vehicle may decrypt the encrypted data using the stored hash chain information. On the other hand, if the vehicle hash chain value is not included in the stored hash chain information, the received broadcast signal may be ignored.

**Lifetime of Secure Cluster**

Secure clusters can define expiration times when initially formed. After this expiration time, the secure cluster can no longer be designed to be valid.

**Secure Clustering Service Using MEC Platform**

Computing offloading may be implemented using an RSU or a base station [7-13]. Secure cluster formation in vehicle communication may optionally be provided. Any vehicle can be provided with secure clustering services using Mobile Edge Computing (MEC) platform, referring to Fig. 6. That is, a vehicle user who desires vehicle communication with improved physical layer security may be provided with a secure clustering service using an MEC platform installed in a device such as a road side unit (RSU) at any time. Here, the secure clustering service may form a hash chain composed of vehicles having a secrecy capacity of a vehicle greater than or equal to a reference value in response to a secure clustering service request executed in an application of the host vehicle, and transmit corresponding hash chain information to the host vehicle. The host vehicle may receive or transmit driving related information to a vehicle included in the hash chain based on the hash chain information.

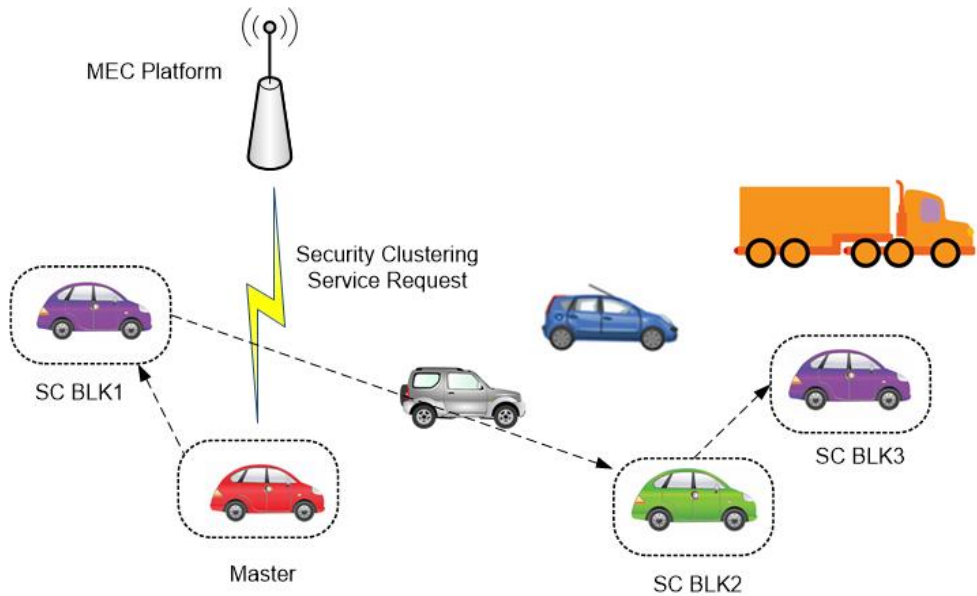

Figure 6: Secure Clustering Service using MEC Platform.

**Conclusion**

We proposed vehicle communications using hash chain-based secure cluster. Each vehicle may disclose the hash chain value to an external vehicle when the vehicular secrecy capacity is greater than or equal to the reference value. MEC platform can use this published vehicle hash chain value to form a vehicle hash chain and send it to each vehicle. As a result, each vehicle can be provided with secure clustering service using the MEC platform to expect more secure vehicle communication. Our research is expected to be essential for the future of autonomous driving and help to apply safe vehicle communication easily.